\documentclass[10pt,aps,twocolumn,prl,showpacs]{revtex4-1}
\usepackage{graphicx} 
\usepackage{amssymb,amsfonts,amsmath}
\usepackage{color}
\newcommand{\schmchn}{%
\begin{figure}[t]
  \begin{center}
   \includegraphics[clip]{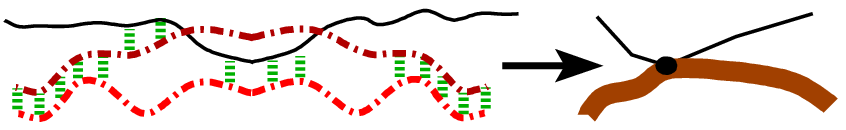}
   \caption {(Color online) Schematic diagram of strand exchange and
     the equivalent coarse-grained three chain interaction.  A single
     strand (solid black line) pairs with the strands (dash-dot lines
     (left)) of a bubble on a duplex (thick line (right)).  Short
     green dotted lines indicate base pairings.  }\label{fig:4}
 \end{center}
\end{figure} }
\newcommand{\bdblock}{
\begin{figure}[htbp]
  \begin{center}
   \includegraphics[clip]{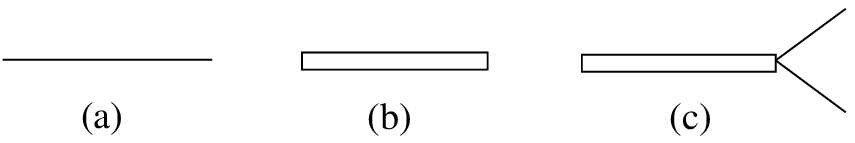}
   \caption{Basic building blocks. (a) represents $Z({\bf k},s)$ for a
     Gaussian chain, (b) $Z_{\sf b}({\bf k},s)$ for a two chain bound
     state,(c) a Y-fork representing the interface between a bound
     pair and two open strands. It has a weight $g_2$.}
   \label{bdblock1}
 \end{center}
\end{figure} }
\newcommand{\bdblockk}{
\begin{figure}[hb]
  \begin{center}
   \includegraphics[clip]{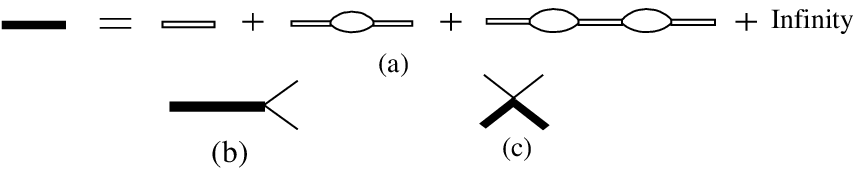}
   \caption{ (a) The duplex partition function as an infinite series
     of bound pairs and bubbles, (b) Y-fork for a duplex. (c) A three
     chain interaction, $g_3$, involving a free-chain and a duplex.
   }\label{bdblock2}
 \end{center}
\end{figure} }
\newcommand{\thchint}{
\begin{figure}[t]
  \begin{center}
   \includegraphics[clip]{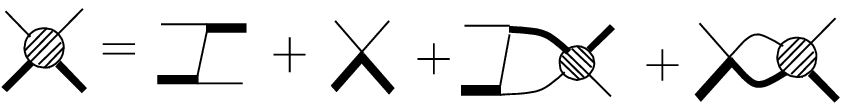}
   \caption{Diagrammatic representation of the three-chain partition
     function. This translates into an integral equation involving
     interactions to all order.}\label{th-ch-int}
 \end{center}
\end{figure} }
\newcommand{\fengy}{
\begin{figure}[t]
  \begin{center}
   \includegraphics[clip]{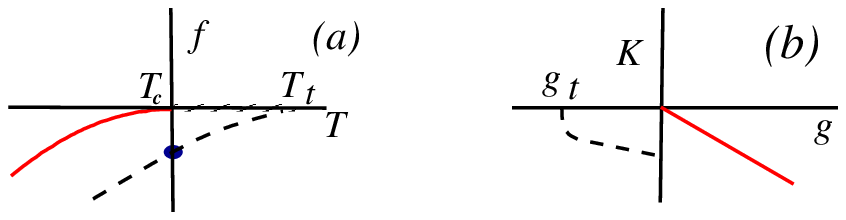}
   \caption{ (Color online) (a) Schematic plot of $f$ {\it vs.} $T$.
     The solid red (dashed black) line is the free energy curve for
     the duplex (triplex) measured from the unbound state.  The
     continuous melting is at $T=T_c$. The filled circle is the
     triplex bound state at $T_c$. The triplex melting at $T_t$ is
     first order.  (b) The corresponding Efimov plot in the $K$-$g$
     plane with line types as in (a).  }\label{freefig}
 \end{center}
\end{figure} }

\newcommand{\pd}[2]{\frac{\partial #1}{\partial #2}}

\begin{document} 

\title{Renormalization Group Limit Cycle for Three-Stranded DNA} 
\author{Tanmoy Pal} \email{tanmoyp@iopb.res.in}
\author{Poulomi Sadhukhan} \email{poulomi@iopb.res.in}
\author{Somendra M Bhattacharjee} \email{somen@iopb.res.in}

\affiliation{Institute of Physics, Bhubaneswar - 751005, India.}

\begin{abstract} 
  We show that there exists an Efimov-like three strand DNA bound
  state at the duplex melting point and it is described by a
  renormalization group limit cycle.  A nonperturbative RG is used to
  obtain this result in a model involving short range pairing only.
  Our results suggest that Efimov physics can be tested in polymeric
  systems.
\end{abstract}

\pacs{87.14.gk, 64.60.ae, 87.15.Zg}
\maketitle

Consider a three-particle quantum system with pairwise short-range
potential.  Apart from the occurrence of the usual three-body bound
state, a very special phenomenon occurs at the critical two-body
zero-energy state. An infinite number of three-body bound states
appear though the corresponding potential is not appropriate to bind
any two of them; the removal of any one of them destroys the bound
state.  This phenomenon, valid for any short-range interaction, is
known as the Efimov effect.  The size of the three-body bound states,
or Efimov trimers, is large compared to the potential range, and so it
is a purely quantum effect\cite{efimov}.  Although it was predicted in
the context of nuclear physics\cite{fonseca,brateen}, it has now been
detected in cold atoms\cite{ultcol}.

An ideal DNA consisting of two Gaussian polymers interacting with
native base pairing undergoes a critical melting transition where the
two strands get detached.  Maji {\it et.al.} recently showed that if,
to a double stranded DNA at its melting point, a third strand is
added, the three together would form a bound state instead of
remaining critical\cite{jm}.  The existence of a triplex has further
been verified by real space renormalization group (RG) and transfer
matrix calculations\cite{jm,jayam}.  That this is an Efimov-like
effect can be seen by the imaginary time transformation of the quantum
problem in the path integral formulation.  The paths in quantum
mechanics are identified as Gaussian polymers and the equal time
interaction maps onto the native base pairing.  Such a bound state of a
triple-stranded DNA is called an Efimov DNA.

In both cases, the special effect is due to a long-range attraction
generated by critical fluctuations at the transition point.  For the
DNA case, the large fluctuations in the bubble sizes at the melting
point allow a third strand to form bound segments with the other two.
The power law behavior of the size of a polymer is essential to induce
a $1/R^2$ interaction between any two chains \cite{jm}.

Universal aspects of polymers are well understood in the RG
approach\cite{oono}.  A single chain and many chain solutions are
described by length scale dependent running parameters, which, with
increasing length scales, are expected to reach certain fixed points.
The purpose of this paper is to show that the triple chain bound state
at the duplex melting point is of a different type.  This ``few-chain
problem'' is actually described by a renormalization group ``limit
cycle''\cite{glazek,brateen}.  The appearance of a limit cycle invokes
log periodicity in the corresponding three-body coupling in the
polymer problem. So they break the continuous scale invariance around
the two-body fixed point imposing a discrete scaling symmetry, the
hallmark of the Efimov states.

Another motivation of this paper is to emphasize that a three-chain
polymer model, a three-stranded DNA in particular, by virtue of
mathematical similarities, provides an alternative system for Efimov
physics.  Triplex DNA is known to occur in nature.  The possibility of
recognizing the bound base pairs of a duplex without opening it, by
forming Hoogsteen pairs, has the potentiality of new types of
antibiotics.  In addition, H-DNA is a common motif formed during many
DNA activities where there is a stretch of triplex DNA via a strand
exchange mechanism\cite{frank}. The advantage with our model is that
the parameters corresponding to the polymers are easily tunable by
changing temperature, solvent conditions, etc.  Some important
phenomena are getting verified in the mathematically analogous low
energy condensed matter systems, e. g., Majorana
fermions\cite{fu-kane}, the Klein paradox\cite{klein}, and structure
formations in early Universe\cite{kibble}.  We hope that this work
will inspire experimental searches of the Efimov effect in polymers.

Because of different critical dimensions of the two-chain and the three-chain 
interactions, namely $d=2$ and $d=1$ respectively, one has to
tackle irrelevant variables in three dimensions, which is outside the
scope of perturbative RG. This makes the problem difficult to handle
in traditional Edwards Hamiltonian\cite{oono} approach.

Our model consists of three directed polymer chains in $(3+1)$
dimensions, where the monomers live in three spatial dimensions (${\bf
  r}$) and we assign an extra dimension along the contour of a polymer
(z).  Two polymers can interact only when they are at the same space
and length coordinate (native base pairing). To avoid difficulties in
solving the full three chain problem, we consider a simplified model
that any two of the three chains are in a bound state.  At zero
temperature, the two chains form a rigid rodlike bound state without
any bending.  At finite temperature, thermal fluctuation melts locally
the bound state to form a number of bubbles, or segments of free-chain
pair, creating interfaces.  A weight factor $g_2$ is given to each
interface or Y-fork. The bubbles allow the paired bound state to bend.
Any unbound chain is taken as free and Gaussian.  A bubble of two free
chains allows a third chain to pair with one of them
(Fig. \ref{fig:4}).  This is the strand exchange mechanism already
alluded to.  Such an exchange generates an effective three-chain
interaction which, in presence of large fluctuations in bubbles,
creates a triplex bound state at the critical point of the two-chain
melting.

\schmchn

Our methodology is to find the two-chain partition function at the
melting point and thereafter
determine the third virial coefficient for  three chains at the duplex
melting point.
Denoting the dimensionless effective three-chain coupling constant by
$H$, our main result shows that the RG beta function is of the form
 \begin{equation}
   \label{eq:5}
   \Lambda\frac{\partial H}{\partial \Lambda}=-A(H-H_0)(H-H_0^*),
 \end{equation} 
where $\Lambda$ is the inverse of a small length scale cutoff, $A$ is
a real constant, and $H_0$ is a complex number.  Because of this pair
of complex fixed points there will be a limit cycle behavior, but most
importantly, there will be a bound state since $H\rightarrow -\infty$.

\bdblock

Fig. \ref{bdblock1} shows the basic building blocks of the model. To
avoid the infinite entropy per unit length of this continuous model,
the unconstrained entropy of a single chain is taken as $\ln\mu$ per
unit length. For the two-chain bound state, we assign an energy
$\propto\epsilon(<0)$ per unit length. We take $k_BT=1$ where $k_B$ is
the Boltzmann factor and $T$ is the temperature. The free chain and
the bound state partition functions for chain length $N$ and
end-to-end distance ${\bf r}$ are given respectively by\cite{dimen}
 \begin{subequations}
  \begin{eqnarray} 
  {\sf Z}({\bf r},N)&=& \mu^{N\Lambda^{2}}(2\pi N)^{-d/2}
                      e^{-\frac{{\bf r^{2}}}{2N}},\\ 
 \hspace*{-14ex}{\rm and}\quad
  {\sf Z}_{\sf b}({\bf r},N)&=&e^{-\epsilon N\Lambda^{2}}
                          (4\pi)^{-1}    \delta({\bf r}-N\Lambda\hat{\bf n}),  
  \end{eqnarray}
 \end{subequations}
where $\hat{\bf n}$ is a unit vector, giving the direction of the rigid rod.  
It is convenient to work in the Fourier-Laplace 
(${\bf k},s$) space where ${\bf k}$ and $s$ are Fourier and Laplace 
conjugates of ${\bf r}$ and $N$ respectively. 
In (${\bf k},s$) space the above partition functions read\cite{nhat}
 \begin{subequations}
  \begin{eqnarray}
    Z({\bf  k},s)&=&(s-\Lambda^{2}\log\mu+k^{2}/2)^{-1},\label{eq:6a}\\
    Z_{\sf b}({\bf k},s)&=&\frac{1}{k\Lambda}\arctan\frac{k\Lambda}
    {s+\epsilon\Lambda^{2}}\stackrel{k\rightarrow0}{=}\frac{1}{s+\epsilon\Lambda^{2}}
    \label{eq:6b}.
  \end{eqnarray}
 \end{subequations}
The Gaussian behavior is reflected in the average size as measured by
the mean squared distance $\langle r^2 \rangle \sim N$.  The poles in
$s$ of Eqs.(\eqref{eq:6a}) and (\eqref{eq:6b}) for $k=0$ at
$\Lambda^2\log\mu$ and $-\epsilon \Lambda^2$ are the negative free
energies of a free chain and a bound pair respectively.  Here ${\bf
  k}=0$ corresponds to the free-end eensemble case. For simplicity,
the $k\to 0$ form of $Z_{\sf b}$ could be used.

\bdblockk

As our partition functions are translationally invariant and a forking
can take place at any $s$, we have ${\bf k},s$ conservation at each
vertex point.  As arbitrary number of bubbles are allowed, the finite
temperature bound state partition function can be written as an
infinite geometric series as shown in Fig. \ref{bdblock2}(a).  We name
the black box on the left-hand side as a {\it duplex}. The duplex partition
function, obtained by summing the infinite series, is given by,
  \begin{equation} 
    Z_{\sf d}({\bf k},s)= Z_{\sf b}({\bf k},s)\left[1-g_{2}^{2}\,I_{0}\,
    Z_{\sf b}({\bf k},s)\right]^{-1}.
  \end{equation} 
Here the single bubble contribution $I_{0}$ is\cite{izero} 
  \begin{equation}
  I_{0}=4\pi\Lambda- 2\pi^2 \sqrt{s'+k^{2}/4}, 
  \label{eq:1}
  \end{equation}
in the limit $(s'+k^{2}/4)\rightarrow0$ where
$s'=s-2\Lambda^{2}\log\mu.$

We want to concentrate on the events near melting where bubbles
proliferate.  For $s{'},k\rightarrow 0$, we find the singularity of
$Z_{\sf d}$ to be
\begin{equation}
  \sqrt{s_*{'}}=-(2\pi^2)^{-1}  g_{2}^{-2} \Lambda^{2} \Delta t , \label{eq:7} 
\end{equation}
where $\Delta t\equiv 2\log\mu+\epsilon -4\pi g_2^2\Lambda^{-1}$ is
the deviation from the duplex melting point with the melting
transition at $\Delta t=0$\cite{deltat}.  The singularity $s_*{'}$ is
the free energy difference of the duplex and two unbound chains.
We choose $g_2$ as the tuning parameter to get melting.  The usual
scaling for ideal DNA melting can also be recovered from
Eq. (\ref{eq:7}), {\it viz.},
 \begin{equation}
    \label{eq:2}
    s_*{'}\sim\xi^{-2}, \ {\rm with}\ \xi\sim |\Delta t|^{-1}.
 \end{equation}
Here, $\xi$ is a diverging length scale for this continuous melting.
In the absence of $g_{2}$ the denaturation transition is first order with
\hbox{$s=-\epsilon\Lambda^{2}$} and $s=2\Lambda^{2}\ln\mu$ for the
bound and the free phases, respectively.

When the system length scale $\xi$ diverges for some critical value of
$g_{2}$ the system goes to the stable free chain phase.  The full
duplex partition function can now be written in terms of $\xi$ in the
small $s{'}$ limit as
\begin{equation} 
    Z_{\sf d}({\bf k},s)=
   \left( {2\pi^{2}g_{2}^{2}\left[-{\xi}^{-1}+\sqrt{s'+
          k^{2}/4}\right]}\right)^{-1}.\label{duplex}
\end{equation}
The two-chain melting behavior is similar to the necklace
model\cite{fisher}.  We note two features.  First, there is a
diverging length scale \hbox{$\xi\sim |\Delta t|^{-1}$}.  Second,
under a scale change ${\bf k}\to b^{-1} {\bf k}$, the length scale and
the free energy change as \hbox{$\xi \to b \xi$} and \hbox{$f\to
  b^{-2} f$}, respectively, for any arbitrary $b$.  This is continuous
scale invariance.

\thchint

Now, let us consider the three-chain case where two of them are in
the duplex state and the third is free.  In the presence of bubbles we
replace Fig. \ref{bdblock1}(c) by Fig. \ref{bdblock2}(b) and call it as
our two-chain coupling $g_{2}$. A duplex can dissociate into two free
chains and if one of them interacts with the third free chain they can
again form a duplex. As a duplex can bend thanks to the presence of
bubbles we can also have diagram like Fig. \ref{bdblock2}(c). This is
our three-chain coupling $g_{3}$.  The free chain-duplex interaction
vertex, $W$, is the third virial coefficient which comes from all the
three-chain connected diagrams as shown in Fig. \ref{th-ch-int}.  The
purpose of introducing the three-chain interaction is to make the
virial coefficient independent of the arbitrary cutoff.  This can be
achieved by doing a momentum shell type integration over a thin shell
which will also give the beta function for the three-chain interaction.

Translational invariance suggests that the partition function depends
only on the duplex-single chain separation at the two end points. The
total momenta at each end can therefore be taken to be zero.  The
evaluation of diagrams of Fig. \ref{th-ch-int} gives
  \begin{eqnarray} 
    W({\bf k,k'},s_{1},s_{1}',s) &=&2g_{2}^{2}Z({\bf
      k+k'},s-s_{1}-s_{1}')+g_{3} \nonumber\\ 
     && \hspace{-.5cm} +(2\pi i)^{-1} \int\, d{\bf q}\;{d\bar{s}} 
                    ( 2g_{2}^{2} {\cal I}_1  +g_{3} {\cal I}_2),
    \label{deval}
  \end{eqnarray}
  \begin{eqnarray}
    &&\hskip -5ex {\rm where,} \quad\quad{\cal I}_1 =
    Z({\bf q},\bar{s})Z({\bf
      k+q},s-s_{1}-\bar{s})\nonumber\\
    &&\qquad \qquad \times Z_{\sf d}(-{\bf q},s-\bar{s})W({\bf q,k'},\bar{s},s_{1}',s),\\
    && \hskip -5ex {\rm and,} \quad\quad{\cal I}_2=
    Z({\bf q},\bar{s})Z_{\sf d}(-{\bf q},s-\bar{s})W({\bf q,k'},\bar{s},s_{1}',s).
 \end{eqnarray} 
We further simplify our model by averaging over all angles and
assuming the external free chains are in relaxed state such that
$(s_1,k)$ and $(s_1',k')$ satisfy the pole condition of Eq.
(\ref{eq:6a})\cite{pole}.  Near melting, $s-3\Lambda^2\log \mu$ is very
small for three chains, and, $\xi\rightarrow\infty$. So most of the
contribution comes from the loop diagrams.  Neglecting the tree level
contributions\cite{Wk}, we have
 \begin{equation}
   \hskip -.4cm  \overline{W}(k)\!=\!\!
   \frac{8}{\sqrt{3}\pi}\!\!\int_{0}^{\Lambda}\!\!\!\! dq\left[
     \frac{1}{q}\log\frac{k^2+q^2+kq}{k^2+q^2-kq}\!
     +\!2k\frac{H(\Lambda)}{\Lambda^{2}}\right] 
   \! \overline{W}(q),
    \label{master}
 \end{equation}
in terms of the redefined  dimensionless quantities
 \begin{equation}
    \label{eq:3}
   \overline{W}(q)= q W(q), \ {\rm and}\quad
   H(\Lambda)=\Lambda^{2} g_{3} (2g_{2})^{-2},
  \end{equation}
For \hbox{$H=0$}, there is no scale 
in the limit $\Lambda\rightarrow\infty$, and Eq. (\eqref{master}) has a solution
 \begin{equation}
    \label{eq:4}
    \overline{W}(k)=C\cos[s_{0}\log(k/\Lambda_{*})],  
 \end{equation}
where $C$, $\Lambda_{*}$ are constants and $s_{0}=1.5036$.  When $H\neq
0$ and $\Lambda$ is finite we can still use this solution as it
retains its form changing only its constants\cite{klock}.

We now use a momentum shell technique to get the behavior of
$H(\Lambda)$.  We first integrate over a small shell of radius
$\Lambda e^{-dl}$ in Eq. (\eqref{master}) and then rescale back
\hbox{$\Lambda\rightarrow\Lambda e^{dl}$} to get the following differential
equation in the limit \hbox{$k\ll\Lambda$,}
  \begin{equation} 
  \frac{1}{\Lambda}\left[\Lambda\pd{H}{\Lambda}-2H\right]
  \int_{0}^{\Lambda}\overline{W}(q)dq+[1+H]\overline{W}(\Lambda)=0.
  \label{diffeq}
  \end{equation}
Using the  form of $\overline{W}$ of Eq. (\eqref{eq:4}), we have the solution
  \begin{equation} 
  H(\Lambda)=-\frac{\sin\left[s_{0}\log
       \frac{\Lambda}{\Lambda_{*}}-
        \arctan\left(\frac{1}{s_{0}}\right)\right]}{\sin\left[s_{0}\log
       \frac{\Lambda}{\Lambda_{*}}+\arctan\left(\frac{1}{s_{0}}\right)\right]}.
  \label{theh}
  \end{equation}
Eqs. (\eqref{diffeq}) and (\eqref{theh}) can be combined to obtain the RG
flow equation of $H$ as given in Eq. \eqref{eq:5} with 
\begin{equation}
   A= {(1+s_{0}^{2})}/{2}, {\rm \ and \ } H_{0}=(1+is_{0})/(1-is_{0}). \label{3beta} 
\end{equation}
If we define a new coupling constant $\zeta=(H-H_{0})/(H-H^{*}_{0})$,
its flow gives limit cycle trajectories\cite{kolom}.  The
$\Lambda$-independence of $H$ and its log-periodicity for large values
of $\Lambda$ can easily be verified from Eq. (\eqref{theh}). As
$g_{2}=$const at the two-chain critical point, $g_{3}$ satisfies 
Eq. (\eqref{3beta}) and
Eq. (\eqref{theh}) respectively.  At the points
$\Lambda_{n}\sim\Lambda_{*}\left(e^{\frac{\pi}{s}}\right)^{n}$, where
$n$'s are integers, $g_{3}$ runs into negative infinity as $\Lambda$
is changed\cite{lambdan}.  As $g_{3}$ can be interpreted as
three-chain binding energy, we get Efimov states at these points.
When $\Lambda$ is increased, $g_{3}$ continuously decreases to
negative infinity and then jumps to positive infinity in a log
periodic manner.  Every jump corresponds to a winding around the limit
cycle or from one Efimov bound state to other.  These states are
concentrated at the origin and infinite in number.

\fengy

So far as the DNA is concerned, the two-chain melting has a critical
behavior with free energy given by $f\sim -|T-T_c|^2$ for $T\to T_c-$,
measured from the unbound state.  The spatial length scale $\xi$,
coming from the fluctuations in the bubble sizes, diverges as given by
Eq. (\eqref{eq:2}).  However the addition of an extra similar strand,
destroys  the
continuous scale invariance mentioned  below Eq. (\eqref{duplex}).
Instead of three critical pairs, one gets  a fluctuation induced bound
state  with a characteristic length scale  $\Lambda_*$.
By using the quantum path integral to polymer mapping,
the DNA partition function can be written as $ Z(N)\sim \sum_n \exp(-
E_0\Lambda_n^2 N), $ where $E_0$ is the ground state energy determined
by $\Lambda_*$, in units such that $E_0 N\Lambda_*^2$ is dimensionless. We assume
that the coefficients do not depend too sensitively on $n$.  For
$N\to\infty$, the thermodynamics is determined by the ground state
energy $E_0$.  If $f$ and $f_N$ represent the free energy per unit
length in the long length limit and for finite length $N$, then one
sees that
\begin{equation}
f = f_N - N^{-1} \ln\left[ 1- \exp(-aNf_{aN}+Nf_n)\right],
\end{equation}
valid only for $a=\exp(2\pi/s_0)$.

Fig. \ref{freefig} shows the equivalence of the free energy curve and
the conventional Efimov plot in QM. The filled circle at $T_c$ is the
schematic three-chain bound state free energy.  The bound state of
size $\Lambda_*^{-1}$, much larger than the hydrogen bond length would
melt at a higher temperature at $T=T_t$. This free energy curve meets
the unbound curve at a finite slope indicating a first order
transition\cite{jm}. The Efimov DNA is observable in the hatched region
between $T_c$ and $T_t$. The equivalent plot for the Efimov case is
shown in (b).  The ground state energy in QM corresponds to the DNA
free energy so that in the Efimov plot, the y-axis of the wave number
($K$) in QM becomes sign$(f)\, \sqrt{|f|}$ while the x-axis is the
temperature deviation from the melting point written in terms of the
inverse duplex length scale, $g={\rm sign}(T_c-T)\,\xi^{-1}$.  One
recovers the square-root behavior of the trimer energy as it
approaches zero and the duplex curve becoming straight line - familiar
from the Efimov plot.

This paper presents an example of the Efimov effect which is more
amenable to experimental verification. Here we show that there exists
a triplex state at the critical point of duplex melting of DNA. The
derivation uses nonperturbative RG of a model involving a short-range
pairing only. We start from a zero temperature, or, completely bound
two-stranded DNA. At nonzero temperature, bubbles form in the bound
state and a third strand can form a duplex with any one of the
denatured pair or both. The renormalization of the short range pairing
generates an effective three-chain interaction which is
responsible for the three-chain bound state. As a result, at the
critical point of two-chain melting, there exists a three-chain bound
state, but no two-chain bound state. The parameters in the case of
polymers are easily tunable, and, therefore, it will be more helpful
in detecting the Efimov effect experimentally.


\begin{thebibliography}{99}
\bibitem{izero} $I_0$ is given by 
\begin{equation}
I_{0}=\int_{{\bf
    q},\bar{s}}\frac{d\bar{s}}{2\pi i} Z(\frac{{\bf k}}{2}-{\bf
  q},\bar{s}) Z(\frac{{\bf k}}{2}+{\bf
  q},s-\bar{s})d{\bf q}.\nonumber
\end{equation}
which can be evaluated using Eq.(\eqref{eq:6a}) exactly to be
\begin{equation} 
  I_{0}=
  4\pi\left[\Lambda-\sqrt{s'+k^{2}/4}\, \tan^{-1}\left(
{\Lambda/}{\sqrt{s'+k^{2}/4}}\right)\right].\nonumber 
\end{equation} 
Eq.\eqref{eq:1} can be obtained in the proper limit.
\bibitem{lambdan} $\Lambda_n$'s are given by
\begin{equation}
\Lambda_{n}=\Lambda_{*}\left(e^{\frac{\pi}{s_0}}\right)^{n}
\exp\left[\frac{\arctan(s_0)-\frac{\pi}{2}}{s_0}\right].\nonumber
\end{equation}
For the quantum problem,  one then gets the
characteristic Efimov tower $E_{n+1}/E_n= $const,  where $E_n$ is the three
particle energy in the $n$th state\cite{efimov}.
\end{thebibliography}
\end{document}